\documentclass[prl,floatfix,superscriptaddress,twocolumn]{revtex4}
\def\av#1{\left\langle#1\right\rangle}
\usepackage{latexsym}
\usepackage{amstext,amsfonts}
\usepackage{epsfig}

\begin{document}

\title{Inter-similarity between coupled networks}

\author{Roni Parshani}
\affiliation{Minerva Center \& Department of Physics, Bar-Ilan University, Ramat Gan, Israel}
\author{Céline Rozenblat} \affiliation{Institute of Geography, Lausanne, Switzerland}
\affiliation{Institute of Geography, Lausanne, Switzerland}
\author{Daniele Ietri}
\affiliation{Institute of Geography, Lausanne, Switzerland}
\author{César Ducruet}
\affiliation{Institute of Geography, Lausanne, Switzerland}
\author{Shlomo Havlin}
\affiliation{Minerva Center \& Department of Physics, Bar-Ilan University, Ramat Gan, Israel}

\date{\today}

\begin{abstract}
Recent studies have shown that a system composed from several randomly interdependent networks 
is extremely vulnerable to random failure. However, real interdependent networks are usually not randomly interdependent, 
rather a pair of dependent nodes are coupled according to some regularity which we coin inter-similarity.
For example, we study a system composed from an interdependent world wide port network and a world wide airport network 
and show that well connected ports tend to couple with well connected airports.
We introduce two quantities for measuring the level of inter-similarity between networks   
(i) Inter degree-degree correlation (IDDC) (ii) Inter-clustering coefficient (ICC).
We then show both by simulation models and by analyzing the port-airport system that as the networks become more inter-similar 
the system becomes significantly more robust to random failure.
\end{abstract}
\maketitle

Recently, an American Congressional Committee highlighted the intensified risk in an attack on national infrastructures, 
due to the growing interdependencies between different infrastructures \cite{commitee}. However, despite the high significance and relevance 
of the subject, only a few studies on interdependent networks exist and these usually focus 
on the analyses of specific real network data \cite{rosato,rinaldi,laprie,kurant2006a,panzieri}. 
The limited progress is mainly due to the absence of theoretical tools for analyzing interdependent systems.
Very recent studies \cite{buldyrev,parshani} present for the first time a framework 
for studying interdependency between networks and show that such interdependencies significantly increases 
the vulnerability of the networks to random attack. In these studies, the dependencies between the networks are assumed to be completely random, i.e., a randomly selected node from network $A$ is connected and depends on a randomly selected node from network $B$ and vice versa. 
Due to the dependencies an initial failure of even a small fraction of nodes from one network 
can lead to an iterative process of failures that can completely fragment both networks.

However, the restriction of random interdependencies is a strong assumption that usually does not occur in many real interdependent systems.
As a first example consider the two infrastructures that are mentioned both in the committee report \cite{commitee} and in the studies discussed above \cite{rosato,buldyrev,parshani}: The Italian power grid and SCADA communication networks. A power node depends on a communication node for control while a communication node depends on a power node for electricity. It is highly unlikely that a central (high degree) communication node will depend on a small (low degree) power node. Rather, it is much more common that a central communication node depends on a central power station.
Moreover, coupled networks usually also poses some similarity in structure, for instance, an area that is overpopulated is bound to have many power stations as well as many communication nodes. 
Another real example is the world wide port and airport networks that we study in this manuscript. 
We find that well connected ports tend to couple to well connected airports 
therefore supporting our assumption that real interdependent networks are usually not randomly interdependent.     

\begin{figure}[h]
\begin{center}
\epsfig{file=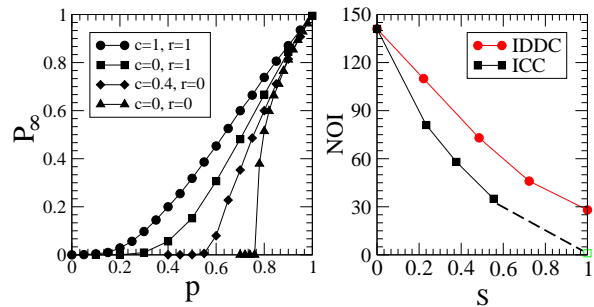,height=4cm,angle=0}
\caption{(a) Simulation results for $P_{\infty}$, the fraction of nodes remaining in the largest cluster of network $B$ 
after a random failure of a fraction $1-p$ of the nodes in network $A$.
The simulations compare four different configurations (see text) of two interdependent SF networks with $\lambda=2.7$ showing that 
inter-similar coupled networks are significantly more robust to random failure compared to randomly interdependent networks (the value of $p$ for which $P_{\infty}$ approaches zero is much smaller).
(b) Simulation results showing the number of iterations (NOI) in the iterative process of cascading failures, at $p_c$. 
The NOI is plotted as a function of the similarity (S) between the networks which is measured either by the IDDC (circles), or by the ICC (squares). 
As the networks become more inter-similar the NOI is reduced indicating that less nodes fail. 
When measuring the effect of the IDDC on the NOI the ICC is kept zero. Similar, when the effect of ICC is measured the IDDC is kept zero. 
The dashed line marks the region that cannot be properly simulated since the ICC is too high to generate networks with no IDDC.}  
\label{compareSimilarity}
\end{center}
\end{figure}

In this Letter we show that inter-similar coupled networks, i.e. coupled networks in which pairs are coupled according to some regularity rather than randomly, 
are significantly more robust to random failure. Moreover, increasing the inter-similarity between the networks leads to a fundamental change in the networks behavior. While randomly interdependent networks disintegrate in a form of a first order phase transition \cite{buldyrev,parshani}, 
networks with high levels of inter-similarity disintegrate in a form of a second order phase transition. 
The phase transition occurs in the size of the largest connected cluster, $P_{\infty}$, of one of the networks (or both) 
when a critical fraction $q_c$ of nodes fail (or a critical fraction $p_c=1-q_c$ remains). For randomly interdependent networks, 
when only a fraction $p_c$ of nodes remains $P_{\infty}$ abruptly drops to zero characterizing a first order phase transition.
For high levels of inter-similarity we find that $P_{\infty}$ continuously decreases at criticality, characterizing a second order transition. 
Fig.~\ref{compareSimilarity} presents simulation results showing the change in the type of phase transition for increasing levels of inter-similarity. 

We develop two measures to asses the level of inter-similarity between interdependent networks. 
We show that these measures can also determine the robustness of coupled networks.
The first quantity, $r^{AB}$, measures the inter degree-degree correlation (IDDC) between a pair of dependent nodes.
The two networks $A$ and $B$ have a degree distribution of $p^A_k$ and $p^B_k$ respectively. 
Similar to assortative mixing in a single network \cite{assortative}, 
we define by $e_{jk}$ the joint probability that a dependency link is connected to an $A$-node with degree $j$ and  
to a $B$-node with degree $k$. For networks with no IDDC, $e_{jk}=p^A_j p^B_k$. For networks with IDDC, 
the level of correlation can be defined by $\sum_{jk}jk(e_{jk}- p^A_j p^B_k)$. 
Normalizing by the maximum value of IDDC we obtain a general measure, $r^{AB}$, in the range $ -1 \leq r \leq 1$. 
The value 1 is achieved for a system with maximum IDDC, the value of zero for no IDDC and a value -1 for a system with maximum anti IDDC.
If the two networks have the same degree distribution ($p_k=p^A_k=p^B_k$) the maximum IDDC value is given by 
$\sigma^2_q = \sum_k k^2p_k - (\sum_k kp_k)^2$ and we obtain
\begin{equation}
r^{AB}= \frac{1}{\sigma^2_q} \sum_{jk}jk(e_{jk}- p_j p_k)
\label{nodeprob}
\end{equation}
Positive values of $r^{AB}$ indicate that high degree nodes from network $A$ tend to couple with high degree nodes from network $B$ and vise versa.
Negative values of $r^{AB}$ indicate that high degree nodes from network $A$ tend to couple with low degree nodes from network $B$ and vise versa.
Randomly interdependent networks correspond to the case of $r^{AB}=0$. 
The second measure is the inter-clustering coefficient (ICC), $c^{AB}$, that evaluates for a pairs of dependent nodes \{$A_j,B_j$\} 
how many of the neighbors of $A_j$ depend on neighbors of $B_j$ and vice versa. 
Analogous to a single network \cite{clustering}, we define the local inter clustering coefficient, $c^A_j$ of node $A_j$ as
\begin{equation}
c^A_j= \frac{t_j}{k^A_j}
\label{nodeprob}
\end{equation}
where $t_j$ is the number of links connecting the neighbors of $A_j$ to the neighbors of $B_j$ and $k^A_j$ is the degree of $A_j$ \cite{note}.
Note that $c^A_j$ is not equal to $c^B_j$. The global ICC can be defined as 
the average of all the local clustering coefficient, $c^A= \frac{1}{N}\sum_j c^A_j$.  
But in this case $c^A \not = c^B$. We therefore prefer to define the global clustering as
\begin{equation}
c^{AB}= \frac{1}{M} \sum_j t_j  
\label{nodeprob}
\end{equation}
where $M$ is the total number of dependency links between the two networks and $0 \leq c^{AB} \leq 1$.
For increasing values of $c^{AB}$ more of the neighbors of $A_i$ depend on the neighbors of $B_i$ and the two networks become more inter-similar. 
For $c^{AB}=1$ the two networks must be identical. 

The effect of inter-similarity between networks is dramatically influenced by the network topology.
In a case where two interdependent networks have a broad degree distribution, an interdependent pair \{$A_j,B_j$\} 
can greatly differ in their degree.
As a result the diversity in the correlation between the networks (IDDC) is significantly increased.       
We therefore apply our theory to two important and very different network topologies. 
The first is the Erd\H{o}s - R\'{e}nyi (ER) network model \cite{ER,ER2,Bollobas}, 
in which all links exist with equal probability leading to a Poisson degree distribution $P(k)=e^{-\av{k}}\av{k}^k/k!$. 
The ER network model has become a classic model in random graph theory and was intensively studied in the past few decades. 
The other model is that of scale free networks (SF) \cite{Barabashi,Barabashi2} networks with a broad degree distribution, usually
in the form of a power-law, $P(k) \sim k^{-\gamma}$ with $\gamma>2$. It was found that many real networks are scale-free \cite{Barabashi,Barabashi2}.
 
To show the effect of our measures on the robustness of inter-dependent networks, we compare between the following four systems:
(i)   Two randomly interdependent SF networks ($r^{AB}=0$ and $c^{AB}=0$).
(ii)  Two SF networks where every pair of dependent nodes \{$A_j$,$B_j$\} has the same degree , $k^A_j=k^B_j$ ($r^{AB}=1$ and $c^{AB}=0$).
(iii) Two interdependent SF networks with $r^{AB}=0$ and $c^{AB}=0.4$, which is the maximum ICC
      we were able to obtain without inserting IDDC.
(iv)  Two identical interdependent SF networks ($r^{AB}=1$ and $c^{AB}=1$).
Fig.\ref{compareSimilarity}(a) presents $P_{\infty}$, the fraction of nodes remaining in the largest cluster of network $B$ 
after a random failure of a fraction $1-p$ of network's $A$ nodes. The Figure shows that high IDDC even with no ICC or high ICC even with no IDDC, significantly increases the fraction of failing nodes (smaller $p_c$) that will fragment the system ($P_{\infty}=0$), 
indicating that the system is more robust. Moreover, for high IDDC or ICC the jump in the size of $P_{\infty}$ that characterizes a first order phase transition changes to a gradual decrease identified with a second order transition.
Fig.\ref{compareSimilarity}(b) provides additional support for our claim that inter-similarity increases the robustness of inter-dependent networks. The number of iterations (NOI) in the process of cascading failures at $p_c$ for a network of size $N$, scales with $N^{1/4}$ for randomly interdependent networks \cite{buldyrev} and is equal to 1 for identical networks. The figure shows that indeed when the inter-similarity is increased either via the IDDC measure or via the ICC measure, the NOI decreases respectively.

Next, we study a real interdependent system composed from the world wide port network and the world wide airport network.
The airport network is composed from 1767 airports and records the majority of the air traffic around the world. 
The port network is composed from 1076 ports and records the flow of commodities around the world.
Previous studies have shown \cite{Capello,Button,Fujita,Rozenblat} that different transportation systems that are located in the same city (or area) depend on each other through their common influence on the economic prosperity of that city. 
In terms of our two networks, the evolvement of an airport in a city will lead to an increase in air-traffic that in tern will result in economic prosperity. The prosperity of that city will have a positive effect on the evolvement and increase of traffic to that city's port and vise versa.   
Accordingly, for our mapping we assume that a port depends on a nearby airport and vise versa.
However, since the networks are not of the same size we first renormalized the networks 
so that they corresponds to the model presented in \cite{buldyrev,parshani}.
We first match between pairs of ports and airports with the minimal distance between them under the condition that a port only depends on one airport and vise versa. The remaining airports that do not depend on any port are merged with the closest airport such that the new renormalized node includes the accumulated traffic of these airports. 
A similar process is applied to the port network. At the end of this process we obtain two networks both of size 992 that are coupled based on geographical location (GL). We find that for the GL interdependent port-airport networks the coupling between the networks is not random, 
the IDDC parameter is $r^{AB}=0.2$ (compared to $r^{AB} \rightarrow 0$ for randomly interdependent networks) indicating that high degree ports tend to couple with high degree airports. These findings supporting our theory regarding inter-similarity between real interdependent networks. 

\begin{figure}[h]
\begin{center}
\epsfig{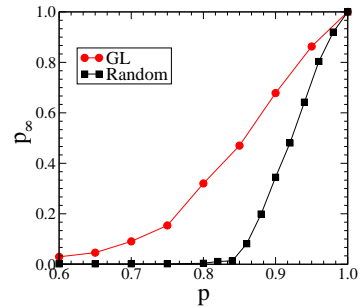}
\caption{The sea-air interdependent system is composed from a world wide port network and a world wide airport network. 
The simulations present the fraction of nodes remaining in the largest cluster of the airport network, $P_{\infty}$, after a fraction $p$ of
the ports in the port network are randomly removed (similar results are obtained for the opposite case). The results are compared between
two different configurations, (i) (Squares) The networks are randomly coupled. (ii) (Circles) The networks are coupled according to geographic locations (GL), i.e., an airport depends on the nearest port and vice versa. 
When the networks are coupled by GL the system is significantly more robust to random failure (the value of $p$ for which $P_{\infty}$ approaches zero is much smaller).} 
\label{compareGLRandom}
\end{center}
\end{figure}

The next step is to enquire how the high level of IDDC in the port-airport networks effects the robustness of the system.
Fig.\ref{compareGLRandom} presents $P_{\infty}$ of the port network for an increasing fraction of failing nodes in the airport network (similar results are obtained when the initial nodes fail from the port network). 
The figure compares between two configurations of the port-airport system: 
(i) The networks are randomly coupled. 
(ii) The networks are coupled based on geographical location (GL).  
The results support our theory that a systems with high IDDC is more robust to random attack ($P_{\infty}$ is larger) and that the phase transition changes from first to second order as the networks become more inter-similar.   
However, since each of the networks has a very high average degree (within each network the nodes are well connected) and as a result the networks are very hard to fragment, we have made the reasonable assumption that even if 75 percent of the traffic to a certain port (or airport) is disabled that port becomes non functional. 

\begin{figure}
\begin{center}
\epsfig{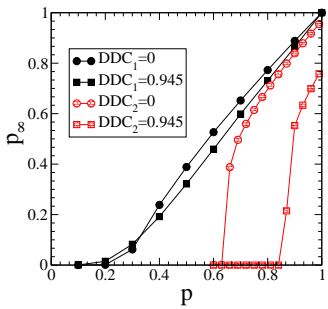}
\caption{Simulation results for ER networks with $\av{k}=3.75$ showing the effect of changing the degree-degree correlation (DDC) 
within a single network (solid symbols) compared to the effect in two interdependent networks (open symbols). 
For the case of interdependent networks the DDC is measured within each network and the IDDC and ICC are kept zero. Even though the effect 
of the DDC on a single network is minor (solid circles) it significantly decreases the robustness of interdependent networks (open circles).} 
\label{single_correlation}
\end{center}
\end{figure}

Until now we have shown the critical effect of inter-similarity on the robustness of a system composed from interdependent networks.
But what is the effect of the local properties within each of the networks on the robustness of the interdependent system?   
Here we show that the degree-degree correlation (DDC) \cite{assortative} within a network that has only a minor effect on the robustness of single networks, greatly effects the robustness of an interdependent system.   
While for single networks a higher DDC usually slightly increases the robustness of the network for an interdependent system
a higher DDC significantly increases the vulnerability of the system.
In Fig~\ref{single_correlation} we demonstrate the effect for the case of ER networks.
For ER networks that are characterized by a very narrow degree distribution, the DDC is expected to have a very limited effect on a single network. But, when two such ER networks with high DDC become randomly interdependent the effect of the DDC becomes dramatic, 
as shown in Fig~\ref{single_correlation}. 

After showing that real coupled networks are indeed inter-similar, we present a mechanism for generating inter-similar coupled networks. 
The model we present can be regarded as a generalization of the Barab\'{a}si-Albert (BA) preferential attachment model \cite{Barabashi,Barabashi2} to two interdependent networks, that naturally incorporates inter degree-degree correlations between the nodes of the two networks. 
According to the BA model, a single network 
with an initial set of $m_0$ randomly connected nodes is grown by adding on each step a new node that is connected to $m$ different nodes from the already existing network. The probability of the new node to connect to a specific node is proportional to that node's degree.    
Generalizing the model to two interdependent networks $A$ and $B$, we start with two initial sets of nodes $m^A_0$ and $m^B_0$ of the same size. 
The two sets are each internally randomly connected and in addition each node from $m^A_0$ is randomly connected to one node in $m^B_0$. 
On each step $t$, a pair of dependent nodes $\{A_t,B_t\}$ are added to the networks, $A_t$ to network $A$ and $B_t$ to network $B$, independently, 
according to the preferential attachment model. Since the two nodes are added independently, there is no correlation between the neighbors of 
node $A_t$ in network $A$ and the neighbors of $B_t$ in network $B$. 
This process mimics a natural process of two interdependent growing network. In terms of our initial example of a power network and a communication 
network, at different times new developing areas are populated and connected to infrastructures. Every such area adds a pair of dependent nodes, 
a power node and a communication node. Even though $A_t$ and $B_t$ are differently connected within each network, 
because of the preferential attachment process the fact that they were added at the same time significantly increases 
the probability that they have a similar degree.
When simulating a system of two interdependent networks according to our generalized BA model we obtained two SF 
networks with $\lambda=3$ as obtained by the BA model for a single SF network \cite{Barabashi,Barabashi2}.
We also obtain a very high level of IDDC ($r^{AB}=0.6$) without any change in the ICC value ($c^{AB}$=0). 
Our model therefore provides a natural mechanism for generating SF inter-similar coupled networks with high inter degree-degree correlation 
but without inter-clustering coefficient. Our simulations also confirm that an interdependent system generated according to the generalized BA model is significantly more robust to random failure than a randomly interdependent system.

We thanks the European EPIWORK project, the Israel Science Foundation, the ONR and the DTRA for financial support.


\begin{thebibliography}{10}
\bibitem{note} According to the assumption made in \cite{buldyrev,parshani} that a node from $A$ depends on at most one node from $B$, 
$k^A_j$ is the maximum number of neighbors of $A_j$ that can depend on neighbors of $B_j$. However, for a more general model where a node can depend on more than one node $k^A_j$ must be replaced by the new maximum.
\end{thebibliography}
\end{document}